# Evolutionary history of the UCP gene family: gene duplication and selection

Joseph Hughes*[1] and Francois Criscuolo[2]

Address: [1]University of Glasgow, IBLS/DEEB, Graham Kerr Building, Glasgow, G12 8QQ, UK and [2]Institut Pluridisciplinaire Hubert Curien, Departement Ecologie, Physiologie et Ethologie, UMR 7178-CNRS, 23 rue Becquerel, 67087 Strasbourg Cedex 2, France

Email: Joseph Hughes* - j.hughes@bio.gla.ac.uk; Francois Criscuolo - francois.criscuolo@c-strasbourg.fr

* Corresponding author

## Abstract

**Background:** The uncoupling protein (UCP) genes belong to the superfamily of electron transport carriers of the mitochondrial inner membrane. Members of the uncoupling protein family are involved in thermogenesis and determining the functional evolution of UCP genes is important to understand the evolution of thermo-regulation in vertebrates.

**Results:** Sequence similarity searches of genome and scaffold data identified homologues of UCP in eutherians, teleosts and the first squamates uncoupling proteins. Phylogenetic analysis was used to characterize the family evolutionary history by identifying two duplications early in vertebrate evolution and two losses in the avian lineage (excluding duplications within a species, excluding the losses due to incompletely sequenced taxa and excluding the losses and duplications inferred through mismatch of species and gene trees). Estimates of synonymous and nonsynonymous substitution rates (dN/dS) and more complex branch and site models suggest that the duplication events were not associated with positive Darwinian selection and that the UCP is constrained by strong purifying selection except for a single site which has undergone positive Darwinian selection, demonstrating that the UCP gene family must be highly conserved.

**Conclusion:** We present a phylogeny describing the evolutionary history of the UCP gene family and show that the genes have evolved through duplications followed by purifying selection except for a single site in the mitochondrial matrix between the 5$^{th}$ and 6$^{th}$ $\alpha$-helices which has undergone positive selection.

## Background

The mitochondrion is the main intracellular site of energy production and is the evolutionary response to the main challenge that living organisms have to face: gaining energy from their environments to sustain their biological functions. The mitochondrial production of ATP is realised by the combination of the phosphorylation of ADP into ATP with an efficient chain of redox reactions, resulting in the so-called oxidative phosphorylation. However, these two processes are not always efficiently coupled, and one reason is the presence in the inner membrane of a family of mitochondrial transporters: the uncoupling proteins (UCP, [1]). UCP1 was first discovered and cloned in 1986 [2] and is involved in the non-shivering thermogenesis (NST) activity of rodent's brown adipose tissue (BAT, [3]). Since then, the discovery of UCP genes has grown rapidly, UCP1 homologues being found across mammalian species (UCP2, UCP3, [4,5]) but also in other eukary-otes from plants to animals [6-8]. Most of the recent attention has been devoted to the evolutionary history of UCP1 since the discovery of UCP1 in ectotherm organisms like teleost fish [9] and amphibians [10]. The fact that organisms that do not show NST possess and express UCP1 raised the question of the exact evolutionary history of UCP1 and of its link with the apparition of thermoregulation. This observation has stimulated an increasing number of phylogenetic studies on UCP [10-15] to determine the origin of the physiological particularity (cold-induced thermogenesis in BAT) in the mammalian lineage [13].

UCP1 and its close homologues (UCP2 and UCP3) are thought to differ in the nature of their uncoupling activity [16,17] and their potential physiological roles (see [18]). Indeed, a rapid overview of the data collected on UCP1, 2 and 3 highlights how these proteins may be different. First, while UCP1 tissue expression is localized (and abundant) to BAT, UCP2 is expressed (in smaller quantities) in a wider range of cell types (like immune or pancreatic β-cells) and UCP3 is mainly present in skeletal muscle ([4,5], see [19]). Also, the physiological role of UCP1 is restricted to thermogenesis, which is unlikely to be the case for UCP2 and 3 as shown by their respective knockout models [20,21]. UCP2 and 3 have been involved in a number of postulated functions in energy regulation, including regulation of insulin secretion [22] or reactive oxygen species production and control of the immune response [20,23,24]. However, accurate data on the mitochondrial activity of UCP2 and UCP3 are still lacking to determine the exact nature of their biological activity [17,25]. Therefore, despite the high sequence identity shared by UCP1, 2 and 3 (close to 60% in humans and mice), punctual amino acid replacement at key structural domains of the respective proteins may have evolved to allow functional specificity to take place. Interestingly, mutagenesis experiments have shown that single amino-acid replacement in UCP1 protein may change its proton permeability (nature of the mitochondrial transport), its sensibility to fatty acid activation or nucleotide inhibition (regulation of the activity, [26]), or its transmembrane structure [27]. The next step in the understanding of the biology of UCP is to determine whether the evolution of UCP genes and protein sequences may have been subjected to different selective pressures after duplication.

Single copy genes are thought to evolve conservatively because of strong negative selective pressure. Gene duplications produce a redundant gene copy and thus release one or both copies from negative selection pressure. There are a number of models for the fate of gene duplicates, the two most prominent of which are neofunctionilization and subfunctionalisation. Thus, duplications are thought to be an important precursor of functional divergence.

The increased availability of UCP sequences in the public databases allows the study of the molecular evolution of the UCP gene family and the evaluation of selection following duplication events. In the present study, we will determine (1) the evolutionary history of the UCP gene family, (2) evaluate the changes in selection pressures following duplications, and (3) identify sites under positive Darwinian selection.

## Results
### *Sequence similarity searches and multiple alignment*
Two lizard sequences from *Anolis carolinensis* were identified during similarity searches with high similarity to UCP2 and UCP3. Homologues of UCP1 were not found in the lizard scaffold genome. Table 1 outlines the sequences (protein and DNA) used in the phylogenetic analyses. It should be noted that additional UCP genes for eutherians and teleosts were identified. Inclusion of these did not improve the reliability of the phylogeny, and as the aim of this study was to determine the evolutionary history of the UCP gene family, only representatives from the major vertebrate clades were included.

### *Phylogeny of the UCP gene family*
The alignments were used to construct phylogenetic trees with maximum likelihood (ML) and Bayesian inference (BI). The different reconstruction methods provided poor support for basal nodes using the protein alignment (Figure 1). The DNA alignment showed support for the UCP1, UCP2 and UCP3 clades, in particular when fewer distantly related outgroups are used perhaps as a consequence of systematic error (Figure 2 and see additional material 1). The different reconstruction methods provided slightly different topologies. Most relationships could be resolved with confidence dividing the gene family into strongly supported clusters in most tree reconstructions. UCP2 and UCP3 genes are sister clusters and the avian UCP gene is grouped within the UCP3 cluster.

The 2 different protein trees were reconciled against a species tree using GeneTree. The protein ML topology required 14 duplications and 47 losses and the BI 15 duplications and 51 losses. The high number of duplications and losses is a result of the basal topology of the gene tree and a number of incongruences between the gene and species trees. However, in the ML protein phylogenies, the basal relationships have low bootstrap supports. Using the DNA phylogeny, the ML tree required less duplications and losses (8 d + 2 l) than the BI tree (12 d + 42 l). The higher number of duplications and losses in the BI reconstruction is mainly a result of duplications inferred through incongruence between the gene and species trees.

**Table 1: List of species and accession numbers for protein and DNA sequences**

| | | Protein | | DNA |
|---|---|---|---|---|
| *Species name* | *Name* | *Accession* | *Name* | *Accession* |
| *Arabidopsis thaliana* | Aratha21593775 | AAM65742.1 | Atha_UCP | NM_115271.4 |
| *Zea mays* | Zeamay19401698 | AAL87666.1 | | |
| *Solanum tuberosum* | | | Stu_UCP | Y11220.1 |
| *Anopheles gambiae* | Anoga11676 | AGAP011676-PA (b) | Aga | XM_552584.3 |
| *Apis mellifera* | Apime66501089 | XP_394267.2 | UCP1Ame | XM_394267 |
| *Strongylocentrotus purpuratus* | Strpur115969038 | XP_001185598.1 | | |
| *Ciona intestinalis* | Cint23999 | ENSCINP00000023999 (b) | UCP_Cint | AK113254.1 |
| *Homo sapiens* | HosapUCP1 | P25874 | UCP1Hsa | NM_021833 |
| *Mus musculus* | MusmuUCP1 | P12242.2 | UCP1Mmu | NM_009463 |
| *Bos taurus* | BotauUCP1 | P10861.2 | UCP1Bta | XM_616977 |
| *Sminthopsis crassicaudata* | SmcraUCP1 | ABR32188.1 | UCP1Scra | EF622232 |
| *Monodelphis domestica* | Modom126331519 | XP_001377555.1 | UCP1Mdol | XM_001377518 |
| *Ornithorhynchus anatinus* | Oana149635652 | XP_001512700.1 | UCP1Oana | XM_001512650 |
| *Xenopus tropicalis* | Xetro166157878 | NP_001107354.1 | Xentrop | NM_001113882.1 |
| *Xenopus laevis* | Xelae147898993 | NP_001088647.1 | UCP1s429 | BC086297 |
| *Danio rerio* | DareUCP4 | NP_955817.1 | UCP4Dre | BC075906 |
| *Danio rerio* | DareUCP3 | AAQ97861.1 | UCP3Dre | AY398428 |
| *Cyprinus carpio* | CypcaUCP1 | AAS10175.2 | UCP1Cca | AY461434 |
| *Tetraodon nigroviridis* | Tenig9630 | ENSTNIP00000009630 (b) | | |
| *Takifugu rubripes* | TakrubUCP1 | ENSTRUP00000033443 (b) | | |
| *Homo sapiens* | HosapUCP2 | P55851.1 | UCP2Hsa | NM_003355 |
| *Mus musculus* | MumusUCP2 | P70406.1 | UCP2Mmu | NM_011671 |
| *Bos taurus* | BotauUCP2 | XP_614452.1 | Bosta_UCP2 | NM_001033611.1 |
| *Antechinus flavipes* | AnflaUCP2 | AAP44414.1 | UCP2Afl | AY233003 |
| *Sminthopsis macroura* | SmmacUCP2 | AAP45779.1 | UCP2Sma | AY232996 |
| *Monodelphis domestica* | ModomUCP2 | XP_001362966.1 | UCP2Mdo | XM_001362929 |
| *Ornithorhynchus anatinus* | OanaUCP2 | XP_001512584.1 | UCP2Oana | XM_001512534 |
| *Anolis carolinensis* (a) | Anca1518 | scaffold_1518:59062–63752 | UCP2_anca | scaffold_1518:59062–63752 |
| *Cyclorana alboguttata* | CycalbUCP2 | ABK96864 | Cycalb_UCP2 | EF065613.1 |
| *Xenopus laevis* | XelaeUCP2 | AAH44682.1 | UCPs1234 | NM_001086754 |
| *Xenopus tropicalis* | XetroUCP2 | AAH63352.1 | UCPxtr | NM_203848 |
| *Cyprinus carpo* | CypcaUCP2 | Q9W725.1 | Cypca_UCP2 | AJ243486.1 |
| *Danio rerio* | DareUCP2 | CAB46268.1 | UCP2Dre | AJ243250 |
| *Tetraodon nigroviridis* | Tetnig47222581 | CAG02946.1 | | |
| *Takifugu rubripes* | TakrubUCP2 | ENSTRUP00000037074 (b) | | |
| *Zoarces viviparus* | ZovivAAT99594 | AAT99594 | | |
| *Homo sapiens* | HosapUCP3 | P55916.1 | UCP3Hsa | NM_003356 |
| *Mus musculus* | MusmuUCP3 | P56501.1 | UCP3Mmu | NM_009464 |
| *Bos taurus* | BotauUCP3 | O77792.1 | UCP3Bta | NM_174210 |
| *Gallus gallus* | GalgaUCP3 | NP_989438.1 | UCPGga | AB088685 |
| *Meleagris gallopavo* | Melgal16755900 | AAL28138 | | |
| *Eupetomena macroura* | Eumac13259162 | AAK16829.1 | UCPEma | AF255729 |
| *Antechinus flavipes* | AnflaUCP3 | AAS45212.1 | UCP3Afl | AY519198 |
| *Monodelphis domestica* | ModomUCP3 | XP_001368096.1 | UCP3Mdo | XM_001368059 |
| *Ornithorhynchus anatinus* | OanaUCP3 | XP_001512822.1 | UCP3Oana | XM_001512772 |
| *Anolis carolinensis* (a) | Anca1149 | scaffold_1149:20424–36291 | Lizard | scaffold_1149:20424–36291 |
| *Xenopus tropicalis* | XentrUCP3 | e_gw1.1014.45.1 * | | |
| *Danio rerio* | Dare50936 | ENSDARP00000050936 (b) | | |
| *Petromyzon marinus* | Pemar51797123 | CO548809.1 | SeaLamprey | CO548809.1 |
| *Lethenteron japonicum* | Lejap149930881 | ABR45662.1 | Letjap_UCP | EF644490.1 |
| *Takifugu rubripes* | TakruUCP3 | ENSTRUP00000037001 (b) | | |

(a) sequence obtained from the February 2007 draft assembly (Broad Institute AnoCar (1.0)) produced by the Broad Institute at MIT and Harvard,
(b) sequences obtained from Ensembl, * sequences from JGI.

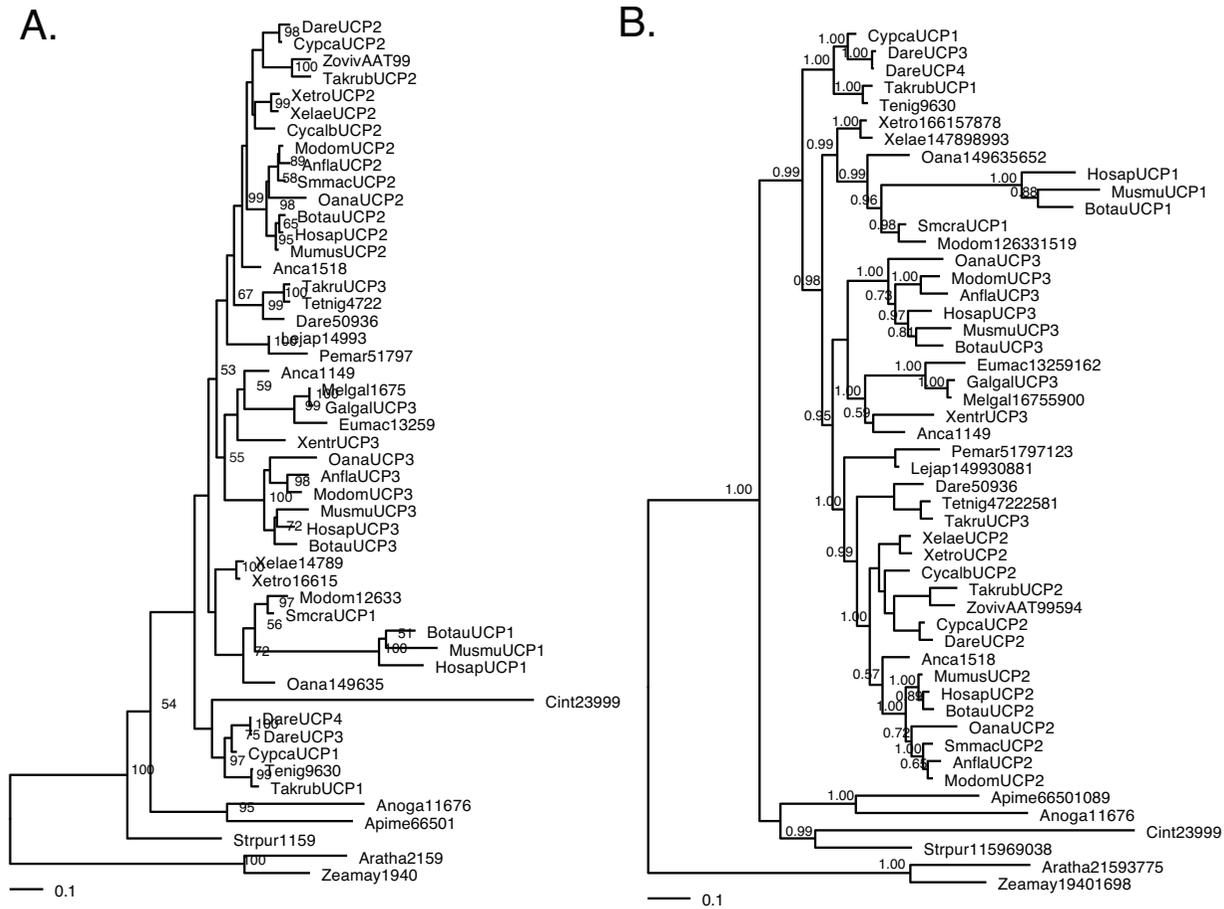

**Figure 1**
**Phylogenetic relationships of proteins within the UCP family**. (A) Maximum likelihood method with bootstrap support (500 pseudo-replicates) above 50% shown at the nodes (likelihood of -7746.18) and (B) Bayesian inference with posterior probability shown at the nodes (likelihood of -8900.97). All trees were rooted with the plant UCP proteins.

If as suggested by the protein phylogenies, the lamprey sequences are sister to the UCP2 clade and the *Takifugu* UCP3 groups within the same clade, then the reconciliation infers 3 duplications (excluding species specific duplications) and 4 losses (excluding losses as a result of incomplete data). However, the results from the DNA phylogenies suggest the lamprey sequences could have diverged before the duplication of UCP2/3. In this case, by removing losses and duplications inferred through mismatch of species and gene trees and losses due to incomplete genome sequences, the most parsimoniously reconciled tree shows 1 zebrafish specific duplication and two major duplications that occurred early in the vertebrate lineage (Fig. 3). One duplication is proposed to have occurred prior to the emergence of teleost fish resulting in two lineages which evolved into UCP1 and UCP2/3 and probably took place early in vertebrate evolution due to the presence of UCP2 in lampreys, although further data are required to confirm the presence of UCP1 and UCP3 in lampreys. A second duplication, also early in vertebrate evolution, resulted in UCP2 and UCP3. Further sequencing of a broader range of ancestral craniata is required to identify a more precise timing for the duplications. Interestingly, UCP2 and UCP1 have been independently lost from the avian lineage but further data are required to confirm the absence of UCP1 in lizards to be able to determine when the loss of UCP1 took place.

### Synonymous and non-synonymous substitution rate estimates
Results using the DNA dataset show that UCP genes are under varying selection pressures (Table 2). Pairwise comparisons of human and mouse orthologs and human and platypus show that UCP1 has higher estimates of dN/dS ratio compared to UCP2 and UCP3 but suggest purifying selection in all three genes. The lower substitution rates for UCP2 and UCP3 shows that they are under strong purifying selection.

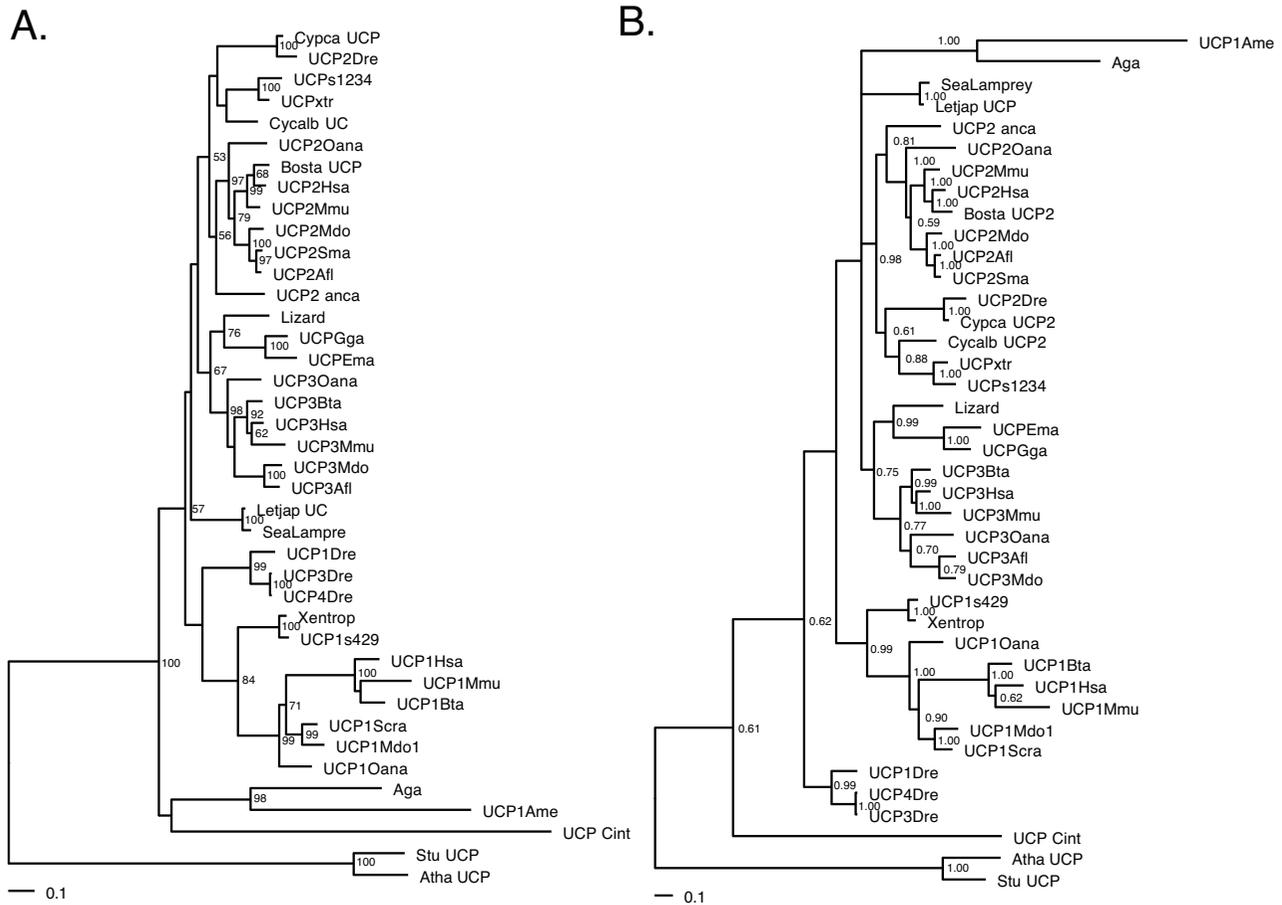

**Figure 2**
**Phylogenetic relationships of DNA sequences**. (A) Maximum likelihood method with bootstrap support (500 pseudo-replicates) above 50% shown at the nodes ((likelihood of -16059.84) and (B) Bayesian inference with posterior probability shown at the nodes (likelihood of -15825.16). All trees were rooted with the plant sequences.

*Positive selection tests*
More sophisticated codon-based substitution models were used to test for branch-specific selection. The model was based on the assumption that selective constraints change following gene duplication. We estimated ω as an average over all sites and branches and the ratio was substantially smaller than 1 (one ratio model ω = 0.07649, Table 3). The one-ratio model was compared with model R2, and the LRT (Table 4) indicated that there is a significant decrease in the rate of non-synonymous substitution following the duplication of UCP1 and UCP2/3 ($\omega_0$ = 0.095 versus $\omega_1$ = 0.066). The comparison of model R2 and R3 also showed that there was a significant difference in the selective pressure following the duplication of UCP2 and UCP3. The branch specific model with three distinct rates of substitution (R3), one for each UCP gene, is a significantly better fit than the one-ratio (R0) and two-ratio (R2) models according to the LRT (Table 4). This suggests significantly different selective pressures on UCP1, UCP2 and UCP3. However, none of the parameters estimated indicate positive Darwinian selection.

The LRT of the one ratio model with M3 indicates that selective pressure is not uniform among sites ($2\delta$ = 919.98, d.f. = 4, p < 0.00001, Table 4). Only the M8 model indicates a site that is evolving under positive Darwinian selection (Table 3). LRTs (Table 4) indicate that model M2 does not fit the data better than M1 whilst it does show that M8 fits the data better than M7, which

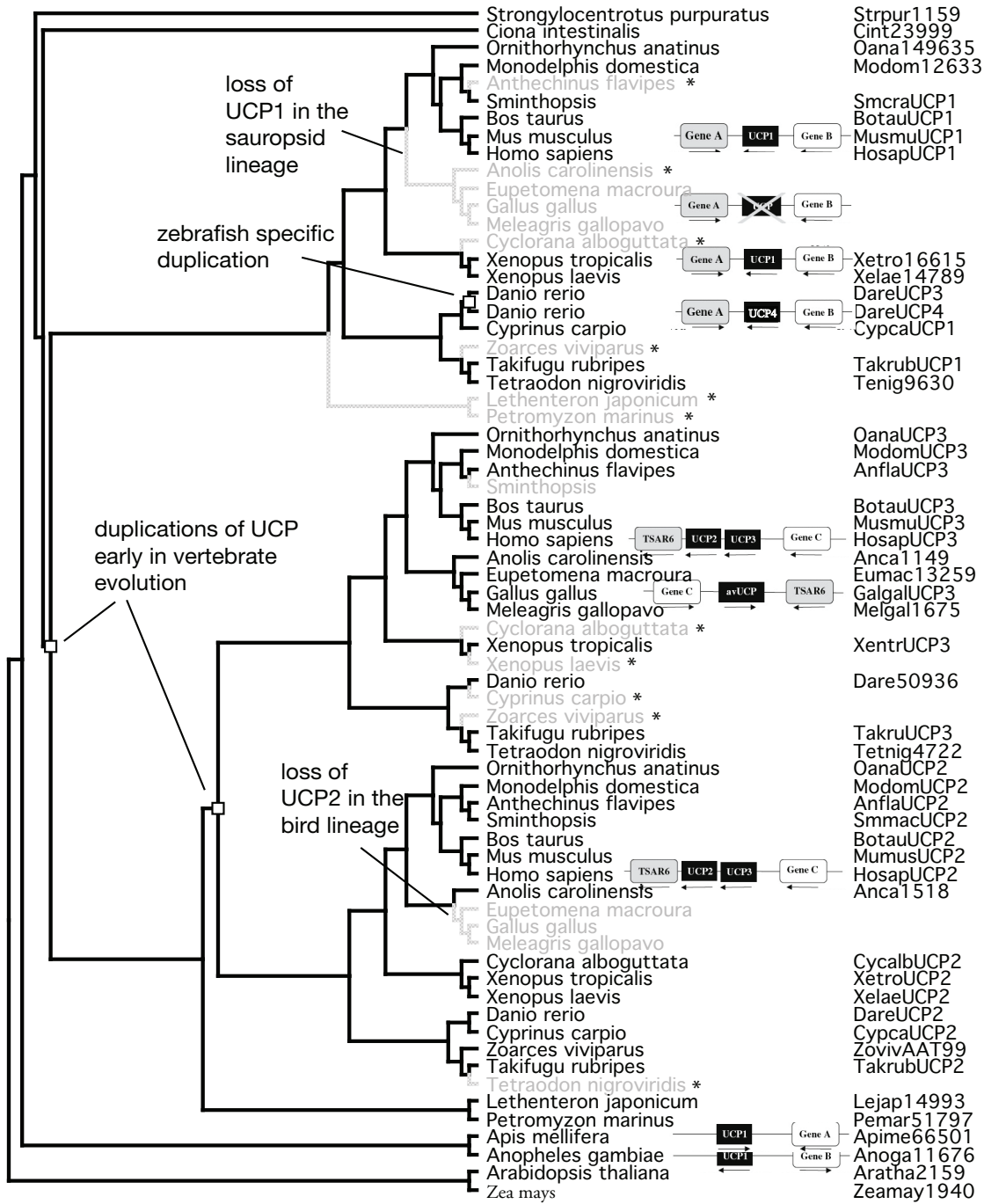

**Figure 3**
**Reconciled tree for the UCP gene family**. The ML tree of UCP genes was reconciled using GeneTree with a species tree. Squares indicate duplication events, grey lines indicate absent genes, either lost from those species or not yet sequenced. An asterisk represents a postulated loss due to incomplete genome sequences. The schematic gene maps of the conserved syntenic regions of the uncoupling proteins according to the study of Emre et al [10].

**Table 2: Synonymous (dS) and nonsynonymous (dN) substitution rates for all UCP genes**

|      | human-mouse | | | human-platypus | | |
| --- | --- | --- | --- | --- | --- | --- |
|      | *dN* | *dS* | *dN/dS* | *dN* | *dS* | *dN/dS* |
| UCP1 | 0.111 | 0.713 | 0.156 | 0.217 | 0.963 | 0.225 |
| UCP2 | 0.018 | 0.672 | 0.027 | 0.044 | 1.083 | 0.041 |
| UCP3 | 0.033 | 0.674 | 0.049 | 0.062 | 1.080 | 0.058 |

Substitution rates were estimated using Yang and Nielson [53] method as implemented in yn00 in the PAML package.

does not allow for positively selected sites (M1 vs M2: $2\delta$ = 0, d.f. = 2, p = 1.0; M7 versus M8: $2\delta$ = 21.12, d.f. = 2, P < 0.0001). Branch-site models were also applied with UCP1 specified as the foreground branch, however the M8 model was the best fit to the data with a likelihood value of – 15233 (Table 3). This model suggests that the variation in selection pressure is due to the evolution by positive selection of a single site, whilst the remaining sites are under strong purifying selection. According to the alignment of the UCP sequences with the 3D structure of the bovine mitochondrial ADP-ATP carrier (PDB id: 2c3e), the site under positive selection is in the mitochondrial matrix between the 5th and 6th alpha-helix (Fig 4). As illustrated using HMM logos [28] for each UCP gene, the site under positive selection follows a highly conserved (Y) amino acid site present across the whole UCP gene family but the site under positive selection is not conserved within the different UCP groups (Fig 5). The HMM logos also illustrate the high level of sequence conservation in the gene family.

## Discussion

In this study, we have sought to expand upon previous phylogenetic studies [10,15,26] by focusing on the UCP gene families and incorporating sequences identified from completed genomes with a subset of cloned sequences, particularly those from non-mammalian species. This study is the first to include lizard UCP genes. The phylogenetic tree reconstruction of DNA sequences gave well resolved topologies with stronger support values for basal relationships than using the protein data probably as a result of the highly conserved protein sequences. These phylogenies provided a method to infer the evolutionary history of the UCP gene family.

The phylogeny of the UCP genes indicates that UCP1, which is present in plants and Arthropods, is the ancestral

**Table 3: Parameter estimates for UCP genes under different branch models, site models and branch-site models**

| Model | Parameters for branches | Positively selected sites | Likelihood |
| --- | --- | --- | --- |
| One-Ratio | $\omega_0$ = 0.07649 | None | -15704.05 |
| *Branch specific* | | | |
| Two-ratios (R2) | $\omega_0$ = 0.0950 | None | -15696.32 |
|  | $\omega_1$ = 0.0660 | | |
| Three-ratios (R3) | $\omega_0$ = 0.0946 | None | -15686.87 |
|  | $\omega_1$ = 0.0845 | | |
|  | $\omega_2$ = 0.0496 | | |
| *Site specific* | | | |
| Neutral (M1) | $\omega_0$ = 0.0657, $\rho_0$ = 0.92529 | Not allowed | -15586.80 |
|  | $\omega_1$ = 1, $\rho_1$ = 0.07471 | | |
| Selection (M2) | $\omega_0$ = 0.06571, $\rho_0$ = 0.9253 |  | -15586.80 |
|  | $\omega_1$ = 1, $\rho_1$ = 0.07470 | | |
|  | $\omega_2$ = **1.266**, $\rho_2$ = 0.00000 | | |
| Discrete (M3) | $\omega_0$ = 0.1109, $\rho_0$ = 0.49068 |  | -15244.05 |
| (K = 3) | $\omega_1$ = 0.10086, $\rho_1$ = 0.38265 | | |
|  | $\omega_2$ = 0.32064, $\rho_2$ = 0.12668 | | |
| Beta (M7) | p = 0.50769 q = 4.86274 | | -15243.99 |
| Beta&$\omega$ (M8) | $\rho_0$ = 0.99, p = 0.53223 | 224 K (P = 0.914) | -15233.43 |
|  | q = 5.57248, $\rho_1$ = 0.00373, $\omega_1$ = **1.69524** | | |
| *Branch-Site* | | | |
| Model A | $\rho_0$ = 0.9155, $\rho_1$ = 0.04732, | In the foreground lineage: | -15563.81 |
|  | $\rho_{2a}$ = 0.03529, $\rho_{2b}$ = 0.00182 | 180 H (P = 0.953), 220 L (P = 0.997), 235 M (P = 0.969) | |
|  | $\omega_2$ = 0.06259 | | |
| Model B | $\rho_0$ = 0.49203, $\rho_1$ = 0.43366, | In foreground lineage: | -15264.48 |
|  | $\rho_{2a}$ = 0.03950, $\rho_{2b}$ = 0.03481 | No significant sites | |
|  | $\omega_0$ = 0.012, $\omega_1$ = 0.12812, $\omega_2$ = 0.52 | In the background lineage: | |
|  | | no significant site | |

The models were implemented in Codeml from PAML. Parameters in bold indicate positive selection.

**Table 4: Likelihood ratio test statistics (2δ) for the test of model fit**

|  | 2δ | df | LRT p |
|---|---|---|---|
| H0 |  | NA | NA |
| One ratio versus H1 | **15.4** | **1** | **<0.001** |
| One ratio versus H2 | **34.3** | **1** | **<0.001** |
| H1 vs H2 | **18.9** | **1** | **<0.001** |
|  |  |  |  |
| LRTs of variable w's among sites |  |  |  |
| One ratio vs. M3 | **919.9** | **4** | **0** |
| M1 vs M2 | 0 | 2 | 1 |
| M7 vs M8 | **21.1** | **2** | **<0.001** |

Significant tests are shown in bold.

UCP as demonstrated in previous studies [10,15,26]. UCP1 then duplicated prior to the divergence of vertebrates. A second duplication of UCP2 and UCP3 also took place early in vertebrate evolution although the exact timing of the event (before or after the divergence of lampreys) requires further genomic data to be gathered. The multiple sequences of UCPs found in the zebrafish, while termed UCP4 and UCP3 are both UCP1 orthologs and should be called UCP1a and UCP1b. UCP4 is syntenic to Gene A and Gene B like other vertebrate UCP1 genes (Figure 3). This could either be a zebrafish specific duplication, or the incomplete sequencing of *Cyprinus carpio* could be hiding an additional paralog and the duplication may be a fish specific genome wide duplications hypothesised to have occurred during fish evolution [29,30]. The latter is probably unlikely due to the lack of duplicates in the complete genome of *Takifugu rubripes*. Importantly, the phylogenetic analyses suggest the independent loss of UCP1 and UCP2 from the avian lineage. The absence of UCP1 in the lizard genome could be attributed to the incompleteness of the genome or could be the result of a loss of UCP1 in the sauropsid lineage.

UCP1 is the only uncoupling protein for which there is a scientific consensus concerning the nature of its physiological function (thermogenesis, [31]). The UCP1 knockout mice are able to maintain their body temperature, but suffer in pronounced cold exposure suggesting that UCP1 is principally involved in short-term adaptation to cold (Enerback et al. 1997). This adaptive evolution probably occurred after the divergence between eutherians and marsupials [13] consistent with the fact that BAT is only found in eutherians. Even though birds are lacking UCP1, they are still able to respond to thermal challenges. The loss of UCP1 and disappearance of BAT in birds is likely due to the concomitant development of physiological adaptations which have replaced BAT function. As evidence, metabolic rate of birds increases in response to cold and body temperature can be maintained [32]. Indeed, induced uncoupling activity in the mitochondria has been found in the skeletal muscle of cold-acclimated birds [33,34] and more recently the implication of UCP3 (avianUCP) has been suggested [35]. These data lead to two non exclusive conclusions. Firstly, birds have evolved other mechanisms of thermoregulation [25] before or after the loss of UCP1 and BAT (e.g.: futile cycle of $Ca^{2+}$ in bird skeletal muscle or greater adenine nucleotide translocase-catalysed proton conductance, [35,36]). Secondly, a fully demonstrated implication of UCP3 (avianUCP) in skeletal non-shivering thermogenesis in birds would suggest that UCP3, which is not involved in thermoregulation in mammals [21,37], has acquired a new function in birds. In this case, the question is whether avianUCP activity could also compensate for the loss of the ucp2 gene, implicated in mammalian immunity [20] and glucose metabolism [22]. This is an interesting point given the non pathologic high chronic glycemia of birds [38].

The molecular evolution of UCP genes showed that they were under strong purifying selection with a significant change towards stronger purifying selection. UCP1 has the highest dN/dS ratio followed by UCP3 and then UCP2. This strong purifying selection highlights the importance of the function of this highly conserved gene family. Although highly variable regions of the sequence which were difficult to assign as homologous were removed from the analyses, the site models showed that adaptation has appeared at a single site located between the 5th and 6th α-helices. The role of this positively selected site has yet to be determined but the amino acid site (Y) immediately prior to it is highly conserved across mitochondrial carriers as are the transmembrane regions that follow the site. Additionally, Saito et al. [13] found that the two amino acid sites that follow this site are conserved in all eutherian mammal ucp1 genes. Based on studies conducted on UCP1, the region delimited by the 5th and 6th α-helices is close to a site of regulation of UCP1 activity by nucleotides and thus could be implicated in the inhibitory control of UCP1 uncoupling effect [15,26]. This region is also hypothesized to be implicated in the mechanism of transport of protons/free fatty acids [39] in UCP1. However, to date there is a gap in the knowledge of the relationship between amino acid sequence and structure for UCP2 and UCP3, and we are unable to speculate on the particular role of this region in these UCP1 homologues. Unfortunately, shared evolutionary history and molecular selection alone cannot be used as the unique criterion to infer protein function, and the true nature of each UCP gene needs to be determined experimentally and independently. Therefore, this positively selected site may play an important functional role and could represent an interesting target site for future mutagenesis experiment thus facilitating our understanding of the structure-function relationships in UCP genes.

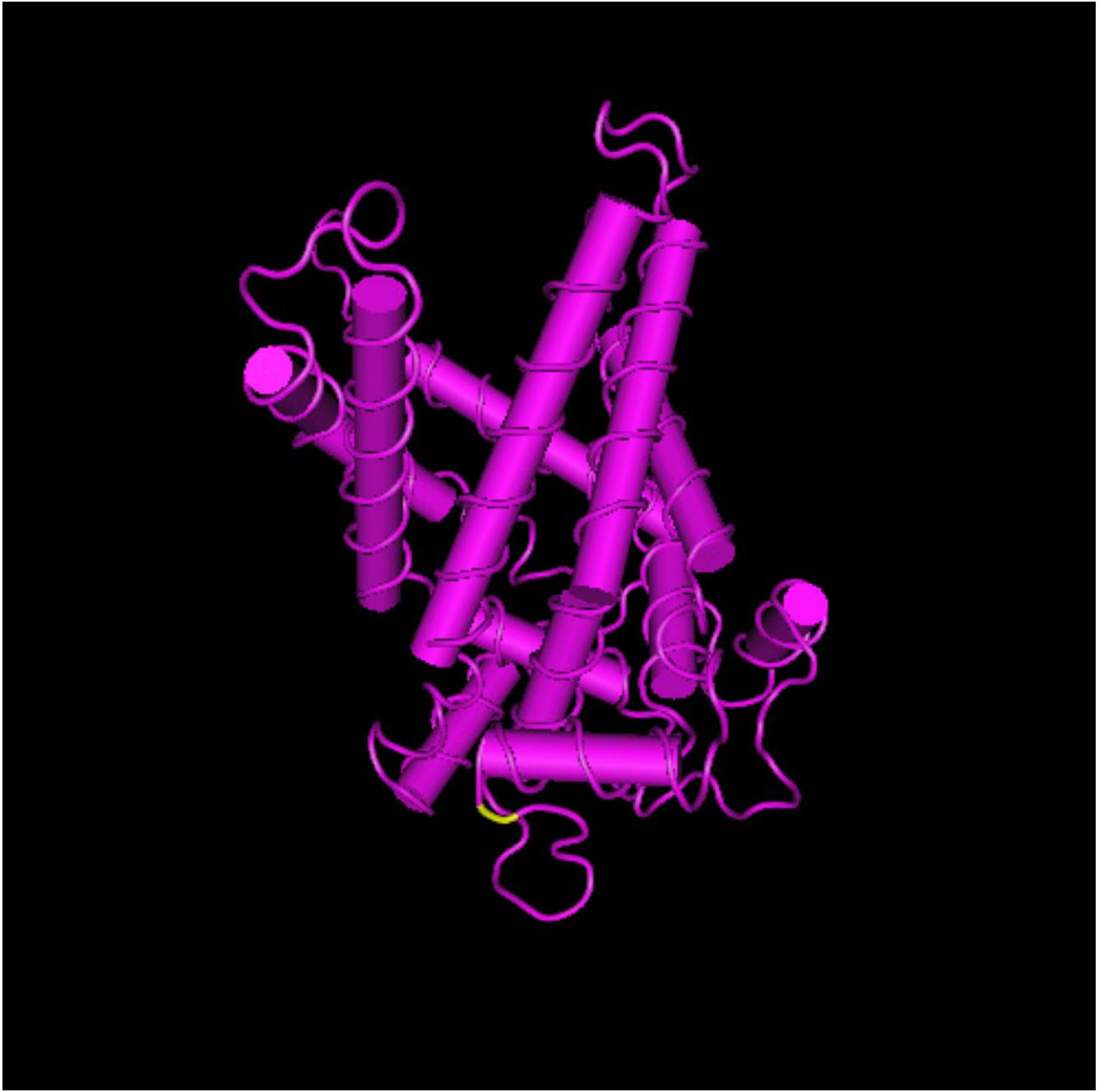

**Figure 4**
**Three-dimensional representation of the UCP molecule according to the 3D structure of bovine mitochondrial adp-atp carrier** (PDB id: 2c3e). The site under positive selection between the 5th and 6th α-helices is shown in yellow.

## Conclusion

Genomic data have provided an opportunity to gain a better understanding about the evolution of UCPs using phylogenetic analyses. The UCP gene family phylogeny shows that two duplications took place early in the evolution of vertebrates. Subsequent to these two duplications, UCP1 and UCP2 were lost from the avian lineage independently. However, further genome projects on a greater diversity of evolutionary lineages are required to better understand the gene-duplication history. Evolutionary rate analysis shows purifying selection across branches and sites (except for one single site with site specific positive selection) suggesting that the function of the genes in the UCP gene family has been highly conserved after

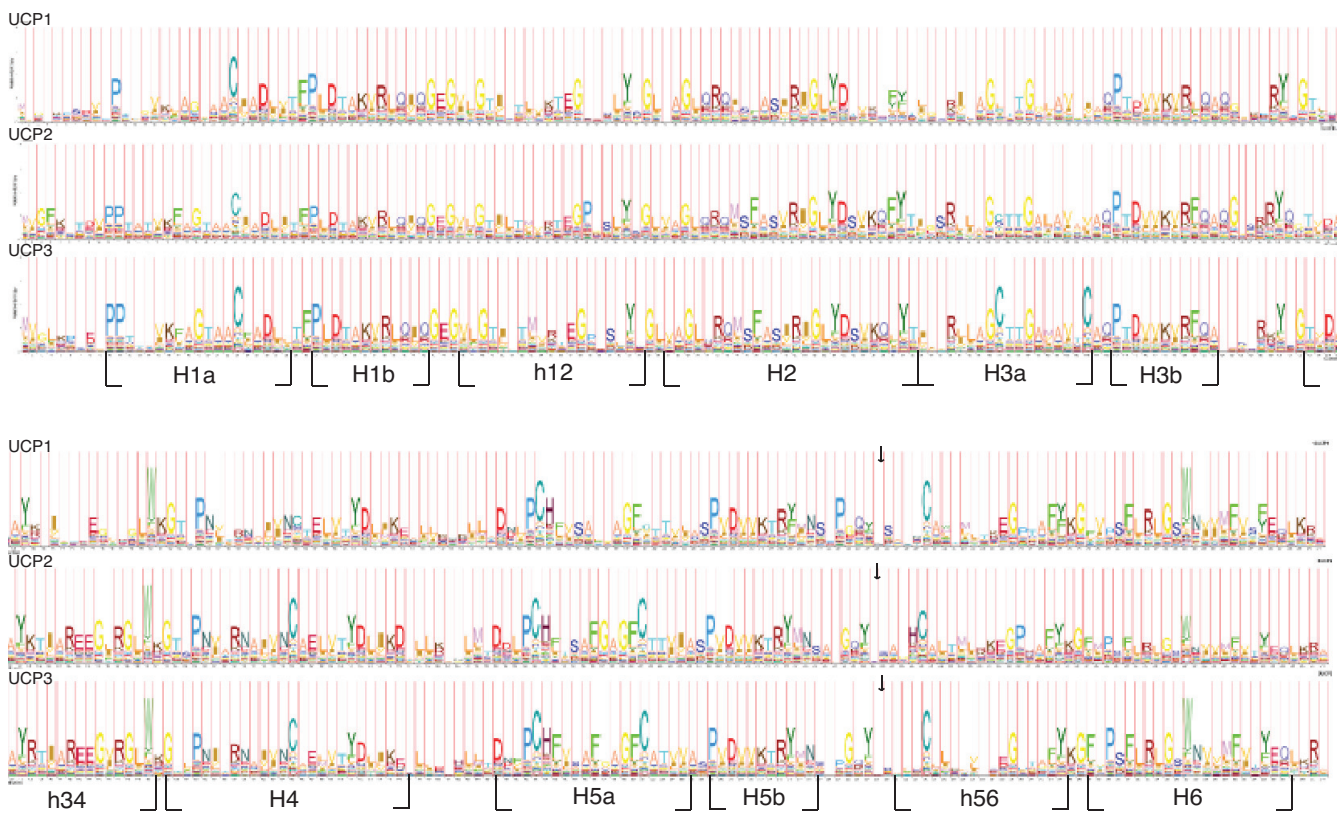

#### Figure 5
**HMM Logos for UCP1, UCP2 and UCP3**. Comparison of the HMM Logos of UCP1, UCP2 and UCP3 protein alignments (excluding variable regions). The numbering of each α-helix follows the nomenclature used for the ADP/ATP carrier [55]. The site under positive selection in model M8 is indicated with an arrow.

duplication events and over evolutionary time. By considering the evolutionary history of the UCP gene family we provide insight into which amino acid residues might have undergone positive selection and could be targeted for site-directed mutagenesis. However, the identification of a single site under positive selection requires supporting evidence from further studies with better algorithms for a more credible assessment of site-specific subfamily divergence.

## Methods
### Sequences and sequence similarity searches
Amino acid and nucleotide sequences of UCP gene family members were obtained from GenBank for most species (see Table 1 for accession numbers). The sequences for the lizard (*Anolis carolinensis*) were obtained from the February 2007 draft assembly (Broad Institute AnoCar (1.0)) produced by the Broad Institute at MIT and Harvard [40]. A total of 50 sequences for 27 species were used in the final analyses.

### Multiple sequence alignment and phylogenetic analysis
Fifty protein sequences were aligned using MUSCLE [43] and gaps, which are problematic in phylogenetic analysis, were removed using Gblocks 0.91b [44]. The final protein dataset was 274 amino acids long (Additional material 2). Uncoupling proteins from plants, insects, the sea squirt and the sea urchin were included. Nucleotide sequences were aligned using ClustalX [41] with the default parameters followed by manual alignment in Macclade [42] according to the amino acid translation. Regions before the starting codon were excluded from the analysis as well as regions poorly aligned due to uncertain homology (positions from the first nucleotide of the start codon: 64–66, 142–180, 331–375, 469–504, 931 to end). The final dataset was 810 nucleotides long (Additional material 3).

Phylogenetic trees were reconstructed using maximum likelihood (ML) implemented in PHYML and Bayesian inference (BI) in MrBayes. Phyml v2.4.4 [45] was used with the online web server [46] for maximum likelihood analysis using the GTR+I+G substitution DNA model selected with ModelTest [47] and JTT substitution model selected with ModelGenerator for the protein analyses [48]. The robustness of the trees were assessed by bootstrapping (500 pseudoreplicates) with PHYML. Bayesian analyses were conducted using the same model with MrBayes v3.1.2 [49]. Node support was assessed as the posterior probability from two independent runs each with four chains of 200,000 generations (sampled at intervals of 100 generations with a burn-in of 1000 trees).

### Reconciliation of gene and species trees
Gene trees of the UCP gene family were reconciled with a species tree using GeneTree [50]. GeneTree attempts to resolve the incongruence between the gene and species trees by predicting duplications and losses [50]. The species tree was based on the Tree of Life phylogeny [51] and NCBI taxonomy [52]. The reconciled tree was edited to remove losses and duplications inferred due to mismatches of the species and gene trees.

### Estimation of substitution rates and testing positive selection
Synonymous (dS) and non-synonymous (dN) substitution rates were estimated using the methods of Yang and Nielson [53] as implemented in yn00 in the PAML software [54]. The two trees (ML and BI) were tested separately for positive selection. Using Codeml from PAML the branch specific models, One-ratio (R1) and Two-ratios (R2) were used to detect lineage-specific changes in selective pressure after the duplication events. The site specific models, Neutral (M1), Selection (M2), Discrete (M3) with 3 site classes, Beta (M7) and Beta&ω (M8) were also used to test for individual residues under positive selection. Likelihood ratio tests (LRT) were used to assess their goodness of fit, by comparing a model that does allow for dN/dS>1 against a model that does not (i.e. null model). Therefore, the branch specific LRT was R2 vs R1. The site specific LRTs were M3, M2 and M8 against their respective null models, M0, M1 and M7. Positively selected sites were listed. Because some of the models like M2 and M8 are noted to be prone to the problem of multiple local optima, we ran the program twice, once with a starting omega value <1 and a second time with a value >1. We used the results corresponding to the highest likelihood.

### Authors' contributions
JH performed all sequence and phylogenetic analysis and drafted the methods and result section of the manuscript, FC conceived the study, participated in the design and coordination of the study and drafted the introduction. Both authors drafted the discussion and read and approved the final version.

## Additional material

**Additional file 1**
*Phylogenies using DNA sequences*. Phylogenies using 29 DNA sequences. (A) Maximum parsimony phylogeny (data analysed in PAUP bootstrapped 1000 times) with bootstrap support above 50 shown at the nodes (tree length of 2763), (B) Maximum likelihood method with bootstrap support (likelihood of -11,992) and (C) Bayesian inference with posterior probability shown at the nodes (likelihood of -11,776). All trees were rooted with Apis mellifera.
Click here for file
[http://www.biomedcentral.com/content/supplementary/1471-2148-8-306-S1.rar]

**Additional file 2**
*Gblocks results*. Sequence alignment of UCP proteins with the selected positions underlined in blue.
Click here for file
[http://www.biomedcentral.com/content/supplementary/1471-2148-8-306-S2.html]

**Additional file 3**
*Nexus matrix*. DNA sequence alignment of UCP genes used for building the phylogenies in the nexus format.
Click here for file
[http://www.biomedcentral.com/content/supplementary/1471-2148-8-306-S3.rar]

## Acknowledgements
The authors wish to thank Barbara Mable for her suggestions and recommendations for the molecular selection analyses. We thank two anonymous reviewers for their comments on this manuscript. This work was conducted during the NERC grant NE/B000079/1 to JH.